# Closed-loop control of gamma oscillations in the brain connections through the transcranial stimulations

Xuan Zhang[1], Duoyu Feng[1], Djibrina Barry[1], Jiajia Li [1, *]

[1]) College of Information and Control Engineering, Xi'an University of Architecture and Technology, Shaanxi, Xi'an, 710055, China;
[*] Correspondence should be addressed to lijiajia_dynamics@xauat.edu.cn

**ABSTRACT**

The reconstruction of brain neural network connections occurs not only during the infancy and early childhood stages of brain development, but also in patients with cognitive impairment in middle and old age under the therapy with stimulated external interference, such as the non-invasive repetitive transcranial magnetic stimulation (rTMS) and the transcranial direct current stimulation(tDCS). However, until now, it is not clear how brain stimulation triggers and controls the reconstruction of neural network connections in the brain.

This paper combines the EEG data analysis and the cortical neuronal network modeling methods. On one hand, an E-I balanced cortical neural network model was constructed under a long-lasting external stimulation of sinusoidal-exponential form TMS or square-wave tDCS was introduced into the network model for simulate the treatment process for the brain connections. On the other hand, by combining Butterworth filter and functional connectivity algorithm, the paper analyzes the relations between the attentional gamma oscillation responses and the brain connection based on the publicly available EEGs during the pre-tDCS and post-tDCS treatment phases.

Firstly, the simulation results indicate that, during long-lasting external stimulations of tDCS/rTMS, The sustained gamma oscillation was found to trigger more release of BDNF from astrocytes to participate in the positively reshaping the excitatory neuronal network connection.

Meanwhile, the simulation results show that after both tDCS/rTMS stimulations, the transient gamma oscillation in response to post-stimulation attentional stimulus inputs become significantly higher than that of before stimulations, accompanying with the improvement of the neuronal network connections. As follows, similar results were found based on the publicly EEG data that the gamma oscillation in response to attentional cognitive task and the functional connectional connectivity were both improved after the tDCS stimulation for the subjects.

The paper adopted the combinations of modeling and EEG data analysis algorithms to provide evince to the conclusion that sustained gamma oscillations during extremal brain stimulations plays a trigger role in the neuronal network connection remodeling in a closed-loop circuit with glial BDNF releasing process, unfolding the intrinsic mechanism of how the non-invasive stimulation improve the brain in a self-repair pathway with the participation of glial cells. This could be ignition for the design of self-optimization for artificial neural networks.

## 1 Introduction

Recent decades of neuroscience research have highlighted the roles of the connectivity patterns of cortical neural networks in the brain functions such as working memory and sensory perception [1-

[3]. The frontiers in this field like the Human Connectome Project open the doors to understanding the complex neural intelligence from the perspective of brain connectivity, which still links neural network intelligence to network connection weights[4-7]. Therefore, research on cortical neural network connections provides new insights into understanding cortical function and artificial neural networks.

In the past decade of connectomics field, the issue of network connection reconstruction, originally termed as metaplasticity by W.C. Abraham and M.F. Bear, is considered one of the most popular research fields and its underlying mechanisms are still debatable. The metaplasticity refers to the phenomenon where neural network connections undergo changes due to prior stimuli, and these changes persist after the termination of the stimuli [8-10]. Studies have utilized this principle to achieve reparative reshaping of abnormal brain states through long-term persistent brain stimulation, such as tDCS, Repetitive Transcranial Magnetic Stimulation (rTMS) and sensory training. For instance, Gold MC found significant cognitive improvement in patients with schizophrenia and depression after 40Hz rTMS[11-15]. Other research utilized this method to improve the brain cognitive performance for the patients of Alzheimer's disease [16-17]. Other Non-intrusive brain stimulations like tDCS[34] and sensory training[18] still were tried into therapy of cognitive impairment synchrome in the patients of traumatic brain injury (TBI)  de Pins B explained this issue of connection re-shaping in the micro spatial scale of the mouse model of Alzheimer Disease, and observed an increase in dendritic spine density induced by the sustained prior stimulation[19] changes in synaptic structures of mice brain subjected to stimulation. Furthermore, various neurophysiological results indicate that gamma oscillations are always associated with the process of network connection reshaping[20]. These results suggest that the role of cortical gamma oscillations is crucial not only in carrying cognitive processes such as cortical information processing, perception, and attention, but also in the process of network connection reshaping induced by sustained prior stimulations. However, the underlying mechanisms of how such network connection reshaping processes are regulated by cortical gamma oscillations remain unrevealed, posing a challenging scientific question.

In the last two decades, the proposal of "Tripartite Synapse" theory has provided new insights into our understanding of the relationship between gamma oscillations and network connection reshaping[19,21-23]. For example, Ameroso and his colleagues found that glial cells, when subjected to gamma oscillations, release BDNF onto the dendrites of postsynaptic neurons to promote the dendritic spine density. This synaptic structural change enhances the information integration capacity of postsynaptic neurons and the process of network connection reshaping [24-26]. Therefore, accurately characterizing the dynamic correlations between gamma oscillations and glial BDNF release emerges critical for revealing the intrinsic mechanisms of the network connection reshaping process in sustained prior stimulations, providing significant theoretical basis for the development of non-invasive stimulation methods.

To achieve the goal, this study first utilized a dynamic modeling approach including mathematical dynamical model of the gamma-glial cell-BDNF loop. It characterized the sustained rTMS and tDCS stimulation as two comparable sustained prior stimulations to explore the dynamic regulatory mechanisms of sustained prior stimulations on the process of neural network reshaping. Subsequently, a public physiological EEG data with tDCS were analyzed by combinations of the complex network and the temporal-frequency algorithms, validating the dynamic regulatory role of gamma oscillations in the process of network connection reshaping highlighted in the modeling part.

## 2 Model and Methods

### 2.1 Cortical Excitation-Inhibition Balance Neural Network Model

In order to explain the mechanism of neural network connection reshaping under sustained stimulation, we propose a cortical neural network composed of feedforward and feedback control loops as shown in **Figure 1**. Here, sustained stimulation first acts on the cortical neural network, triggering cluster gamma rhythm rapid discharge, inducing nearby glial cells to release BDNF, and further driving the update of neural network connections. The update of neural network connections will feedback to the cortical neural network, leading to the improvement of the response effect of the neural network in the attention testing section. As shown in the long-term stimulation phase in the middle of **Figure 1**, two system inputs, periodical sinusoidal from of rTMS and sustained square wave of tDCS, were used. Among them, short-range attention square wave stimulation was used as the test stimulation, combined with the conclusions of this study and previous research[51], by detecting the level of system gamma oscillation response to evaluate the cognitive effect of attention.

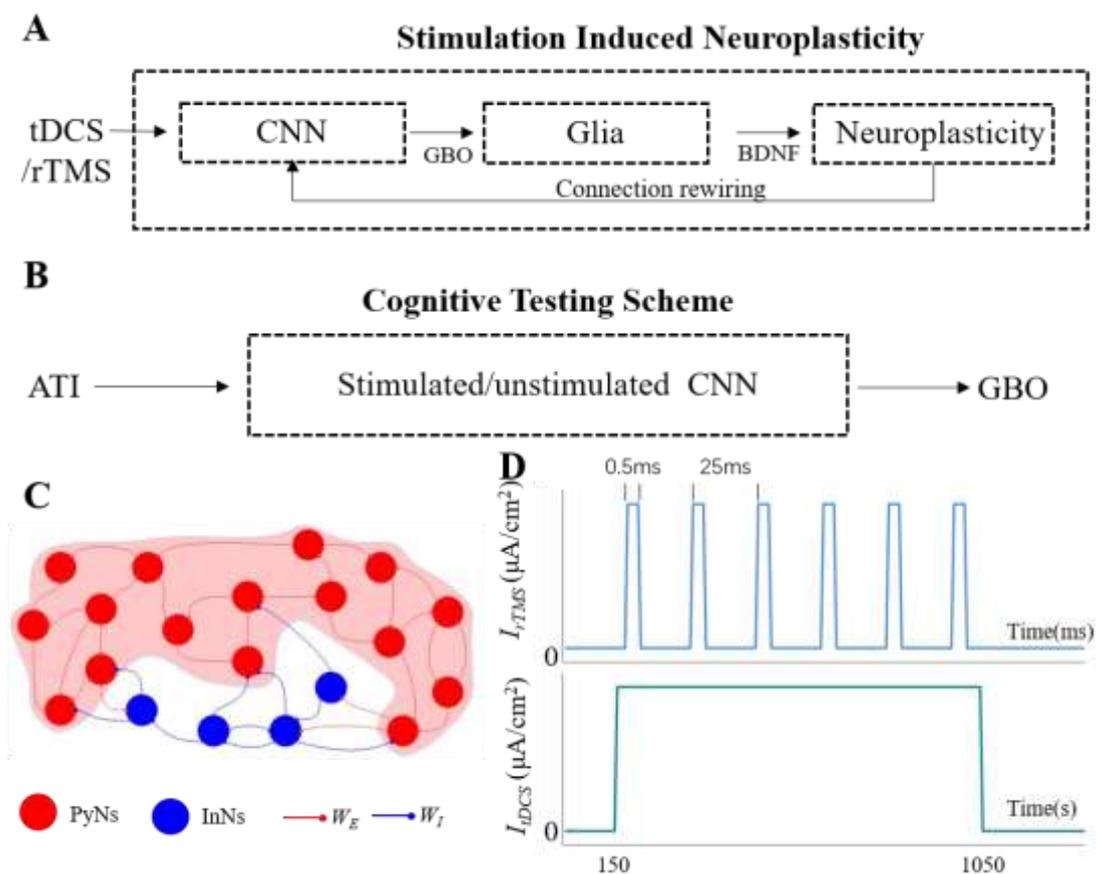

**Figure 1** Model Scheme of cortical neuronal network imposed with the transcranial stimulations of tDCS/rTMS. A depicts the model structure, consisting of pyramidal neurons (Excitatory neuron, E), interneurons (I), and glial cells (Glia). The E-I neuron cluster forms the Cognitive Neural Network (CNN), generating gamma oscillation (GBO) discharges. These discharges influence glial cells through ATP, leading to the release of neurotrophic substance (BDNF) into the synaptic cleft. B illustrates model of cognitive testing scheme, involving the attentional stimulus input to the CNN to test the state of simulated codex. C presents a schematic diagram of the CNN model, comprising pyramidal neurons, interneurons, and glial cells. D illustrates the application of stimuli, simulating tDCS/rTMS lasting for

15 minutes.

At the microscale, the specific form of the E-I balanced neuronal network compromising with 400 excitatory pyramidal cells (E) and 100 inhibitory interneurons (I) mentioned in **Figure 1**A is illustrated in **Figure 1C**. The inter-population connectivity model in the target neuron population is implemented in a mean-field manner, which is dependent on the mean firing rate of the source neuron population. The considered system was also termed as a PING (pyramidal-interneuron network gamma) neural network model, simulating the gamma-frequency (30-100Hz) periodic neuronal spiking during cognitive tasks, as shown in **Figure 1C**. The model considers the dendritic part of pyramidal neurons receiving external stimuli and serving as the primary location for carrying excitatory electroencephalographic propagation. Wang's two-compartment model for cortical pyramidal neurons was adopted[31][31], considering a 1:4 ratio of real cortical interneuron groups to pyramidal neuron groups. Spike discharge influences glial cells through neurotransmitter ATP, causing them to release BDNF into the synaptic clefts enveloping the neurons. The BDNF released into the gap stimulates neurons on both sides of the synaptic cleft, enhancing their connection strength, thereby promoting an increase in the connection strength of the entire E-I neuron cluster and ultimately enhancing its discharge intensity and synchronization.

Considering the slow dynamics of the glial cell system and the overall simulation timeframe for tens of minutes, for computation simplification, we used a linear system transfer function approach to fit the dynamic model of BDNF based on experimental observation data: the BDNF concentration response to the square wave input of neurotransmitter ATP to glial cell[27]. This model takes into account the integration of action potential inputs from nearby pyramidal neurons by glial cells and the reshaping effect of the release of BDNF on the receptor density of dendritic spines in post-synaptic neurons[28].

During the long-lasting therapied tDCS/rTMS phase, we apply a 15-minute tDCS/rTMS to the model to simulate transcranial stimulation, reflecting the stimulation parameters in real experiments. Only under such a timescale of stimulation can the brain's connection structure undergo remodeling. We simulate the changes in connection strength caused by the elevation of BDNF concentration during the long-lasting stimulation phase by modifying the internal connection strength of pyramidal neurons in the first layer. In the cognitive testing phase, before and after modifying the connection strength of pyramidal neurons by tDCS/rTMS, we apply a short-duration attentional stimulus input (ATI) for 10s to the model. The presence of glial cells mainly integrates neurotransmitter inputs induced by action potentials from nearby pyramidal neurons and releases BDNF to post-synaptic neurons, reshaping the receptor density of dendritic spines in pyramidal neurons[28].

It should be noted that the equivalent input to the dendrites of pyramidal neurons under transcranial stimulation follows the description in the model section, referring to the mathematical descriptions of such simulations as described in the literatures[29-30], as shown in **Figure 1D**.

Pyramidal neurons are modeled using Wang's two-compartment model[31], which involves an isolated dendrite model that better describes the integration process of dendritic synaptic inputs from other neurons in the network. The model is as follows:

$$C_m \frac{dV_{E,i}}{dt} = -I_L - I_{Na} - I_K - I_{Ca} - I_{AHP} - \frac{g_c}{p}\left(V_{E,i} - V_{d,i}\right) \qquad (1)$$

$$C_m \frac{dV_{d,i}}{dt} = -I_L - I_{Ca} - I_{AHP} - \frac{g_c}{(1-p)}(V_{d,i} - V_{E,i}) + I_{syn} \quad (2)$$

Where $C_m$ represents the neuron membrane capacitance, action potentials of the soma and dendrites in the neuron model are respectively represented by $V_E$ and $V_d$. The current between the soma and dendrites is proportional to $(V_E-V_d)$, with units of μA/cm². $g_c$ represents the coupling conductance, and p represents the ratio of soma cell area to total area, indicating the difference in current intensity due to the different membrane surface areas in the cell. The soma of each neuron consists of leakage current $I_L$, sodium ion current $I_{Na}$, potassium ion current $I_K$, calcium ion current $I_{Ca}$, and afterhyperpolarization current $I_{AHP}$. The dendrites of each neuron consist of leakage current $I_L$, calcium ion current $I_{Ca}$, and afterhyperpolarization current $I_{AHP}$. The interaction of these currents can generate intrinsic rhythmic firing patterns in individual neurons. Their mathematical descriptions are as follows:

$$I_L = g_L(V - V_L) \quad (3)$$

$$I_{Na} = g_{Na} m_\infty^3(V) h(V - V_{Na}) \quad (4)$$

$$I_K = g_K n^4(V - V_K) \quad (5)$$

$$I_{Ca} = g_{Ca} m_\infty(V - V_{Ca}) \quad (6)$$

Whereas, voltage-dependent currents are described in a form similar to the HH model, so the gating variables m, h, and n satisfy:

$$m_\infty = \alpha_m / (\alpha_m + \beta_m)$$
$$\alpha_m = -0.1(V+33)/\{\exp[-0.1(V+33)]-1\} \quad (7)$$
$$\beta_m = 4\exp[-(V+58)/12]$$

$$\frac{dh}{dt} = \phi_h[\alpha_h(1-h) - \beta_h h]$$
$$\alpha_h = 0.07\exp[-(V+50)/10] \quad (8)$$
$$\beta_h = 1/\{\exp[-0.1(V+20)]+1\}$$

$$\frac{dn}{dt} = \phi_x[\alpha_n(1-n) - \beta_n n]$$
$$\alpha_n = -0.01(V+34)/\{\exp[-0.1(V+34)]-1\} \quad (9)$$
$$\beta_n = 0.125\exp[-(V+44)/25]$$

The assumption is that the intracellular calcium concentration is controlled by leak integration, so the voltage-independent calcium-activated potassium current $I_{AHP}$ (afterhyperpolarization current) takes the form of:

$$I_{AHP} = g_{AHP}\left[[Ca^{2+}]/([Ca^{2+}]+K_D)\right]((V-V_K)) \tag{10}$$

$$\frac{d[Ca^{2+}]}{dt} = -\alpha I_{Ca} - [Ca^{2+}]/\tau_{Ca} \tag{11}$$

Where α is proportional to S/V (membrane surface area to volume inside the membrane), we use α = 0.002, with units of μM(ms·μA)$^{-1}$cm$^2$, and the calcium influx for each spike is 200nM. Various pumping and buffering mechanisms are collectively described by a first-order decay process with a time constant $\tau_{Ca}$ = 80ms. Pyramidal neurons are connected to nearby pyramidal neurons, and pyramidal neurons are connected to interneurons in a mean filed form which is dependent on the mean firing rate of interneuron population. Their mathematical descriptions are as follows:

$$I_{syn} = I_{Vd}^{E \to E} + I_{Vi}^{I \to E} \tag{12}$$

$$I_{V_I}^{I \to E} = g_{GABA}\overline{V_I}\sum_{j=1}^{N_I} y_{GABA,j}(E_{GABA}-V_{E,i}) \tag{13}$$

$$\frac{d y_{GABA}}{dt} = \alpha_g(1-y_{GABA}) - 0.18 y_{GABA} \tag{14}$$

$$\alpha_g = 5/[1+\exp(-V/2)] \tag{15}$$

$$I_{Vd}^{E \to E} = g_{AMPA}\overline{y_{AMPA}}(E_{AMPA}-V_{d,i}) \tag{16}$$

$$\frac{d y_{AMPA}}{dt} = \text{Heaviside}(V_{E,i}+10) - y_{AMPA}/2 \tag{17}$$

Where $g_{GABA}$ represents the synaptic conductance of GABA synapses, $y_{GABA}$ represents the open-close state of GABA channels, $g_{AMPA}$ represents the synaptic conductance of AMPA synapses, and $y_{AMPA}$ represents the opening state of AMPA channels. The Heaviside step function, also known as the "unit step function," can be represented by returning whether x exceeds zero to denote a piecewise constant function. The parameters in the equation are shown in the table below

Table 1 Parameter Sheet of excitatory pyramidal cell model

| Parameters | Values | Parameters | Values |
| --- | --- | --- | --- |
| $g_c$ | 2.0ms/cm$^2$ | $E_{ca}$ | 120mV |
| $p$ | 0.5 | $g_L$ | 0.1ms/cm$^2$ |
| $C_m$ | 1.0μF/cm$^2$ | $g_{Na}$ | 45ms/cm$^2$ |
| $g_{GABA}$ | 0.1 | $g_K$ | 18ms/cm$^2$ |
| $g_{inh}$ | 0.2 | $g_{Ca}$ | 1ms/cm$^2$ |
| $g_{AMPA}$ | 0.05 | $g_{AHP}$ | 0/5ms/cm$^2$ |
| $g_{autapse\_in}$ | 0.1 | $K_D$ | 30 |
| $E_{AMPA}$ | 0mV | $\alpha$ | 0.002 |
| $E_{GABAa}$ | -75mV | $t_{Ca}$ | 1 |
| $E_L$ | -65mV | $\Phi_h$ | 4 |
| $E_{Na}$ | 55mV | $\Phi_n$ | 4 |
| $E_K$ | -80mV | | |

For simplifying the computation, fast-spiking interneurons adopt the model established by Wang et al. for fast-spiking interneurons[32]. Theoretical studies indicate that when suitable conditions for synaptic transmission are met, these GABAergic interconnections can synchronize the neuronal networks in cortex. We utilize the model of fast-spiking interneurons to provide gamma frequency oscillations through GABA$_A$ receptor-mediated synaptic transmission, which is employed for synchronizing the discharge of neuron populations across spatial distributions. The equations for fast-spiking interneurons are as follows:

$$C_m \frac{dV_{I,j}}{dt} = -I_{Na} - I_K - I_L + I_{syn} \tag{18}$$

Where $C_m$ represents the neuron membrane capacitance, the neuron soma is composed of sodium, potassium, and leakage currents. The external synaptic stimuli to the neuron include inhibitory currents brought by nearby interneurons, and excitatory currents brought by pyramidal neurons. The sodium and potassium currents generating spikes are similar to the HH model, and their mathematical descriptions are as follows:

$$I_L = g_L (V - V_L) \tag{19}$$

$$I_{Na} = g_{Na} m_\infty^3 (V) h (V - V_{Na}) \tag{20}$$

$$I_K = g_K n^4 (V - V_K) \tag{21}$$

The mathematical descriptions of the gating variables are as follows, where m, h, and n represent the opening states of ion channels, with the mathematical expressions as follows:

$$m_\infty = \alpha_m / (\alpha_m + \beta_m) \tag{22}$$

$$\alpha_m = -0.1(V + 35) / \{\exp[-0.1(V + 35)] - 1\} \tag{23}$$

$$\beta_m = 4\exp[-(V + 60)/18] \tag{24}$$

$$\frac{dh}{dt} = \phi_x [\alpha_h (1 - h) - \beta_h h] \tag{25}$$

$$\alpha_h = 0.07\exp[-(V + 58)/20] \tag{26}$$

$$\beta_h = 1 / \{\exp[-0.1(V + 28)] + 1\} \tag{27}$$

$$\frac{dn}{dt} = \phi_x \left[ \alpha_n (1-n) - \beta_n n \right] \tag{28}$$

$$\alpha_n = -0.01(V+34)/\{\exp[-0.1(V+34)]-1\} \tag{29}$$

$$\beta_n = 0.125\exp[-(V+44)/80] \tag{30}$$

These dynamic expressions are modified from the Hodgkin-Huxley (HH) model to capture the characteristics of fast-spiking interneurons in the cortex. The stimulation received by the interneurons is provided by pyramidal neurons, nearby interneurons, and autoinhibitory currents. Their mathematical expressions are as follows:

$$I_{syn} = I_{V_E}^{E \to I} + I_{V_I}^{autapse} + I_{V_I}^{I \to I} \tag{30}$$

$$I_{V_E}^{E \to I} = g_{AMPA} \overline{V_E} \sum_{i=1}^{N_E} y_{AMPA,i}(E_{AMPA} - V_{I,j}) \tag{31}$$

$$I_{V_{I,j}}^{self} = g_{in} y_{GABA}(E_{GABA} - V_{I,j}) \tag{32}$$

$$I_{V_{I,j}}^{I \to I} = g_{inh}(\overline{V_I} - V_{I,j}) \tag{33}$$

Where $g_{GABA}$ represents the synaptic conductance of GABA synapses, $y_{GABA}$ represents the opening state of GABA channels, $g_{AMPA}$ represents the synaptic conductance of AMPA synapses, and $y_{AMPA}$ represents the opening state of AMPA channels. The parameters in the equation are shown in the following table.

Table 2 Parameters sheet of Inhibitory I neurons

| Parameters | Values | Parameters | Values |
| --- | --- | --- | --- |
| $g_L$ | 0.1 | $V_L$ | -65.0 |
| $g_{Na}$ | 35.0 | $V_{Na}$ | 55.0 |
| $g_K$ | 9.0 | $V_K$ | -90.0 |
| $g_{AMPA}$ | 0.05 | $\varphi x$ | 5 |
| $g_{in}$ | 0.1 | $E_{AMPA}$ | 0 |
| $g_{inh}$ | 0.2 | $E_{GABA}$ | -75 |

Finally, most studies of glial cells dynamics models focusing on Glia-neuron interactions aim to reveal the influence of glial cells on synaptic transmission and plasticity. Glial cells express receptors for various neurotransmitters and modulators, including glutamate, ATP, adenosine, and purines, and exhibit calcium signaling in response to their stimulation. Studies have emphasize that While glial cells secrete neurotrophic factor BDNF using glutamate as an exogenous stimulus[33], this may not be limited to this type of neurotransmitter; instead, different neurotransmitters could be expected to trigger similar secretion of internal BDNF. Vignoli and Canossa's study suggests that, besides glutamate, extracellular nucleotide ATP also regulates the secretion of internalized BDNF in cultured cortical glial cells[28].

The focus of our attention on glial cells is the relationship between ATP and BDNF. Based on the research data of Vignoli and Canossa, we constructed a nonlinear system, mathematically described as follows:

$$\frac{[BDNF]}{dt} = k_{ATP} / (1 + \exp(-V_E / 2)) - a_{BDNF} \cdot [BDNF]. \tag{34}$$

$$C_m \frac{dV_d}{dt} = I + g_{AMPA} \cdot (E_{AMPA} - V_d) \cdot \kappa_{BDNF} \tag{35}$$

$$\kappa_{BDNF} = \frac{[BDNF]^n}{k^n + [BDNF]^n} \tag{36}$$

In the equation, [BDNF] denotes the rise in BDNF concentration resulting from action potential modulated ATP release from neurons, the parameters were fit based on the the research data of Vignoli and Canossa[28], where $k_{ATP}$ = 0.00001861 and α$_{BDNF}$ = 0.00000031121. In the equation, $k$ describes a saturation effect on BDNF concentration, with $k$ = 0.18 and $n$ = 1.

**2.2 EEG Data Processing**

The scalp EEG data used for cognitive tasks were obtained from the OpenNeuro dataset[46] regarding cognitive tasks. The dataset employed the international 10-20 system for electrode placement, with a sampling frequency of 512Hz, 64 channels, and 49 subjects. Each subject's experimental protocol consisted of three parts. In the first part, the subjects underwent a cognitive experiment and EEG data were recorded. In the second part, the subjects underwent a 20-minute tDCS stimulation session while EEG data were recorded. Subsequently, the subjects underwent another cognitive task as the third part of the experiment. Each subject repeated the experiment after a one-week interval. Based on the goal of this study on the functional connection and GBO response after the external stimulations, we focused on analyzing the correlation changes in the prefrontal cortex, parietal cortex, and occipital cortex. To extract the required features, we processed the data using filtering methods and directed correlation analysis, shown as follows.

**2.2.1 GBO power analysis algorithm for EEG/simulated neural responses**

In order to explore the hidden characteristics of oscillations in EEG signals a, we employed a filtering method to further process the data. The raw signals were first passed through a band-pass filter with a range of 30-100Hz for further filtering. This processing not only extracts prominent gamma features but also helps reduce computational complexity. Additionally, we further calculated the quadratic form of the filtered time series to obtain equivalent power values. Since the previous step retained the original sampling frequency, the gamma power sequence obtained at this stage contains considerable micro-oscillation noise interference. Therefore, we further smoothed the obtained gamma power sequence using sliding time windows to obtain the final dynamic gamma oscillation power sequence.

**2.2.2　Directly correlation analysis method for EEG data**

Brain, far from merely responding to input stimuli and producing output, which represents the most basic behavior, involves the coordinated efforts of the visual cortex, premotor cortex, amygdala, and hypothalamus. Human behavior entails not just simple reflexes but rather multi-cortical integrated information processing[47]. To understand the regularities in brain network functional connectivity from a data analysis perspective and compare them with the structural connectivity outcomes of model neuron populations, we first employed Pearson correlation analysis to assess the functional connectivity of 64 electrode electroencephalogram (EEG) data. According to experimental findings, both before and after receiving transcranial stimulation therapy, the directly functional connectivity (DFC) between the occipital lobe channels and other channels during visual cognitive tasks better described the cognitive state of the brain at that time. Furthermore, based on the network correlation matrix obtained in the first step, we calculated the average connection strength between individual channels and channels O1, O2, and O3, and other independent channels. It was observed that DFC changes in the frontal and parietal regions were most significant in this regard. As the subjects underwent visual cognitive tests, it is suggested that the DFC was mainly obtained between the primary visual cortex and the other brain regions, as shown in **Figure 2**.

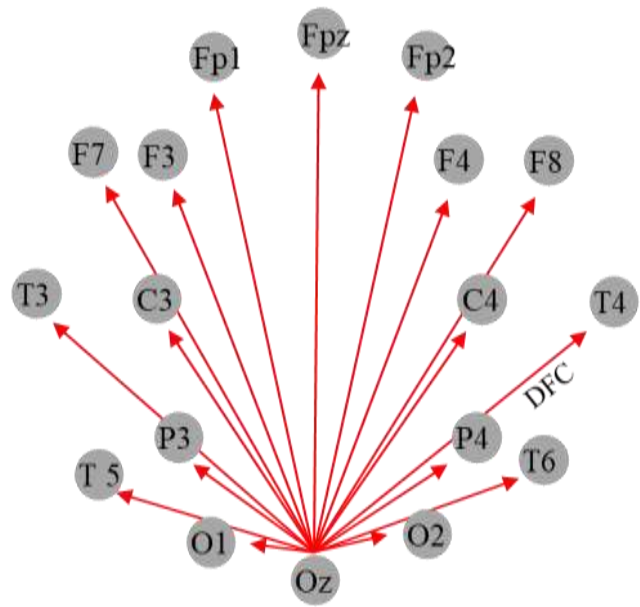

**Figure 2** Directly functional connectivity DFC between primary cortex and the whole brain regions

## 3 Results

Experiments have demonstrated that applying appropriate transcranial stimulation to the brain can enhance cognitive performance, which has a promoting effect on traumatic injuries and neurodegenerative diseases[34]. Also, studies have shown that the promoting effects arise mainly from the increase of neuronal BDNF concentrations induced by those stimulations[35], in accompany with enhanced gamma-band responses in the brain[36]. Based on this, to explain the mechanism of neural network reshaping under sustained transcranial stimulations, we propose the feedforward and feedback control loop shown in **Figure 1**. In this loop, sustained stimulation first acts on cortical neural networks to trigger fast discharge (gamma oscillation), further inducing adjacent glial cells to release BDNF and driving neural network connection updates. The connection updates of the neural network will feed back to the cortical neural network, resulting in improved response effects of the neural network in the attention testing（AT）. The testing algorithm involves short-range attention square wave stimulations, referred in our previous research[20,51], to evaluate the cognitive level of attention based on gamma oscillation responses.

### 3.1 Neuroplasticity and Gamma oscillation under tDCS stimulation

To explore the cortical neural discharge patterns and cortical neural connection reshaping under tDCS firstly, we designed the simulated tDCS stimulation lasting approximately 15 minutes and two ATIs before and after tDCS to be imposed onto the E -I balanced neuronal network explained in **Model and Methods** section. The remaining phases were designed as resting states without stimulations.

Firstly, the neural responses under such stimulation were discussed in the spiking diagrams of excitatory E neurons (blue) and inhibitory I neurons (red) in **Figure 3**. **Figures 4** A-C depict the zoomed detailed diagrams of three phases from **Figure 3**, and their corresponding averaged power spectral distribution. The three phases included the resting phase before the first ATI stimulation, the first ATI stimulation and the second ATI.

From the neural spiking diagrams in **Figure 3,** it can be observed that the neurons exhibit sparse discharge during the resting phase without stimulation (rest phase), while during the ATIs and tDCS stimulation phases, they display dense discharge. Meanwhile, during the initial resting phase the detailed spiking activity in **Figure 4 A** show that the neurons exhibit a chaotic spiking pattern, and their frequency concentrated to the low-frequency domain, as shown in the corresponding power spectrum (**Figure 4** D). However, during the two ATIs before and after tDCS stimulation (Figures 4 B, C), a dense synchronous spiking pattern is observed. Particularly noteworthy is that during the second ATI phase following tDCS stimulation (**Figure 4**C), the synchronization is higher than those during the first ATI (**Figure 4** B). Also, from the corresponding power spectra (**Figure 4** E,**Figure 4** F), it can be seen that the power of the gamma band 30-100Hz during the second ATI (**Figure 4** F) is much improved than that during the first ATI.

Through the comparison of the neurons discharge in the time and frequency domains in **Figure 4**, we can conclude that 15 minutes of tDCS stimulation caused changes in the neuronal connections of the mode , and enhanced the synchronization and gamma-band power distribution of the neuron population spiking discharge.

In previous research, Schwertfeger et al. [34] found that tDCS stimulation can improve cognitive function by altering default network cortex. The work of Reinhart and Nguyen[37] suggests that synchronization of brainwaves in brain cortex can improve working memory in older adults and younger individuals with poor brain function, and synchronization of brain rhythms may improve cognitive abilities in populations with poor cognitive abilities. Also, asynchronous frontal cortical rhythms can impair adaptive control and learning[38]. Therefore, the improved firing synchronization of neurons populations presented in this paper implied a therapied pathway of the tDCS simulation for the poor cognitive abilities in populations.

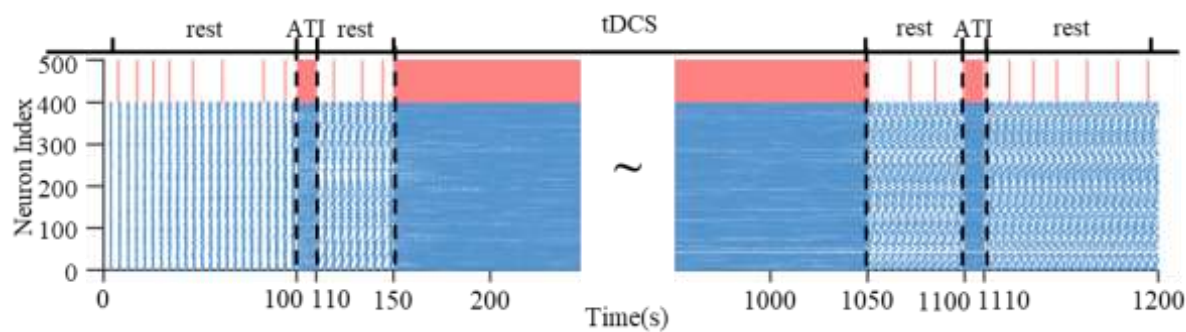

**Figure 3** The spiking diagram of the neuron population discharge inclluding 100 I neurons(red) and

400 E neurons(blue) after receiving external stimulation shown at the bottom. In sequence, the stimulation received: 100 seconds of rest without stimulation, 10 seconds of square wave stimulation to simulate attention stimulation, 40 seconds of rest, 900 seconds of tDCS stimulation to simulate transcranial stimulation, 50 seconds of rest, 10 seconds of square wave stimulation to simulate attention stimulation after receiving transcranial stimulation therapy, and 90 seconds of rest until the end.

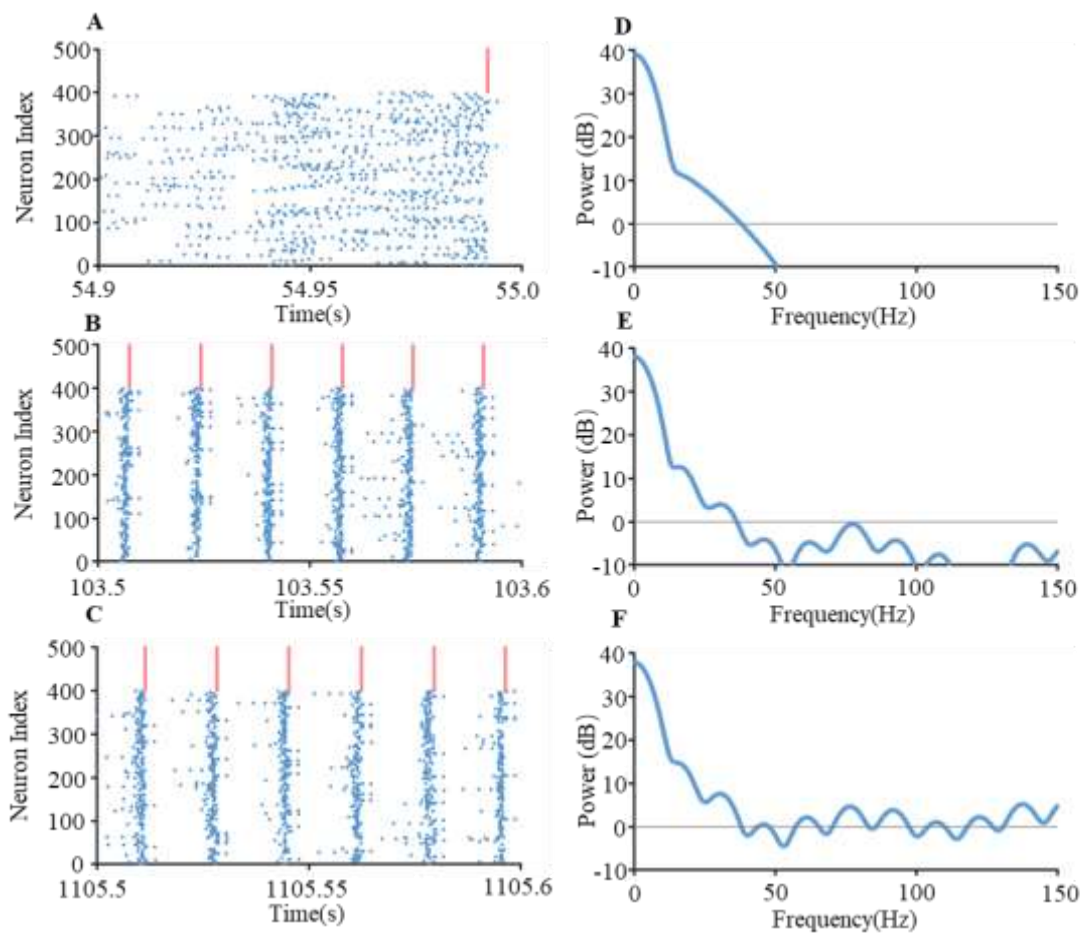

**Figure 4** Temporal and frequency responses of neuron population discharge during stimulation phases selected from **Figure 3**. A-C are detailed spiking diagrams from **Figure 3** time domain: the first rest phase (55.5-55.6 seconds), the first ATI phase (103.5-103.6 seconds), and the second ATI phase (1105.5-1105.6 seconds). **D-F** represent the power spectral diagrams of neuron population discharge during the corresponding phases to the left panel, respectively.

In the above, we mentioned that continuous tDCS stimulation led to changes in neuronal network connection of the, notably in accompany with the improved gamma oscillations. Because fast neuron population firing prompts glial cells to release BDNF. The increased concentration of BDNF alters the connection strength of the neuron populations, and in turn enhances synchronization and discharge frequency of the neuron populations. To comprehensively investigated the roles of BDNF and gamma power in shaping connection strength of the neuron populations, we discussed the dynamical BDNF response, gamma oscillation and the network connection ratio $\kappa_{BDNF}$ modulated by BDNF in **Figure 5**, wherein the dynamical gamma oscillation was obtained by employing filtering and time-frequency analysis methods explained in **Model and Methods** section.
**Figure 5** shows the average power time evolution diagram of gamma power of the neuron populations, along with the corresponding trend of BDNF over time. The gamma oscillation in **Figure 5**A was obtained by the filtered and time-frequency analysis algorithm, with a moving time window of 0.5 seconds and an overlap of 0.25 seconds. From **Figure 5**A, it can be observed that during the rest state, the model hardly produces gamma-frequency neural discharge. However, during the two ATI phases and the tDCS stimulation phase, the neuron population discharge exhibits a higher gamma oscillation amplitude than that in the rest state. Simultaneously, there is a noticeable dynamical consistency between the BDNF concentration and the network connection ratio $\kappa_{BDNF}$ in **Figure 5B-5C**. This indicates that the BDNF concentration is directly proportional to the strength of neuronal connections. It can also be observed that during the 15-minute tDCS stimulation process, the gamma oscillation persisting in a high value is accompanied by a rapid increase in BDNF concentration and $\kappa_{BDNF}$. During the ATI phases, the different level in the BDNF concentration and $\kappa_{BDNF}$ between the two phases result in a significant increase in the gamma oscillation amplitude.

From this result, we can conclude that, on one hand, applying continuous tDCS stimulation to the model enhances gamma oscillations, promoting an increase in BDNF. On the other hand, the increased BDNF in turn drives improvements in structural connections and plasticity. The changes in BDNF concentration reflect the external manifestation of increased E-I network connection strength in the model and an external manifestation of brain plasticity.

Numerous biological experiments have shown that sustained cognitive tDCS simulations benefits the improvement of cognitive levels in patients with conditions such as Alzheimer's disease, accompanied by enhanced network connections after training[39]. Continuous stimulation can induce brain gamma oscillations, which have been proven to be proportional to cognitive levels[40]. Under conditions of sustained stimulation, BDNF significantly increases compared to before stimulation, and BDNF can facilitate structural and functional synaptic plasticity in the hippocampus, cerebral cortex, and many other brain regions[41]. The analysis of the results above indicates that the enhancement of connections is sufficient to induce changes in gamma oscillations during ATI phases. This result aligns with findings from the previous visual experiments, which discovered that an increase in excitatory synaptic strength promotes the generation of gamma oscillations[42].

In conclusion, continuous tDSC stimulation can induce gamma oscillations in the brain, and gamma oscillations, in turn, promote an increase in BDNF in neuronal synapses. The elevated BDNF concentration, in reverse, can enhance structural and functional synaptic plasticity in the cerebral cortex. Through simulating continuous tDCS stimulation, raising gamma oscillations and BDNF concentration, and enhancing network connections, this study from a computational model perspective provides evidence that prolonged stimulation can improve cognitive levels.

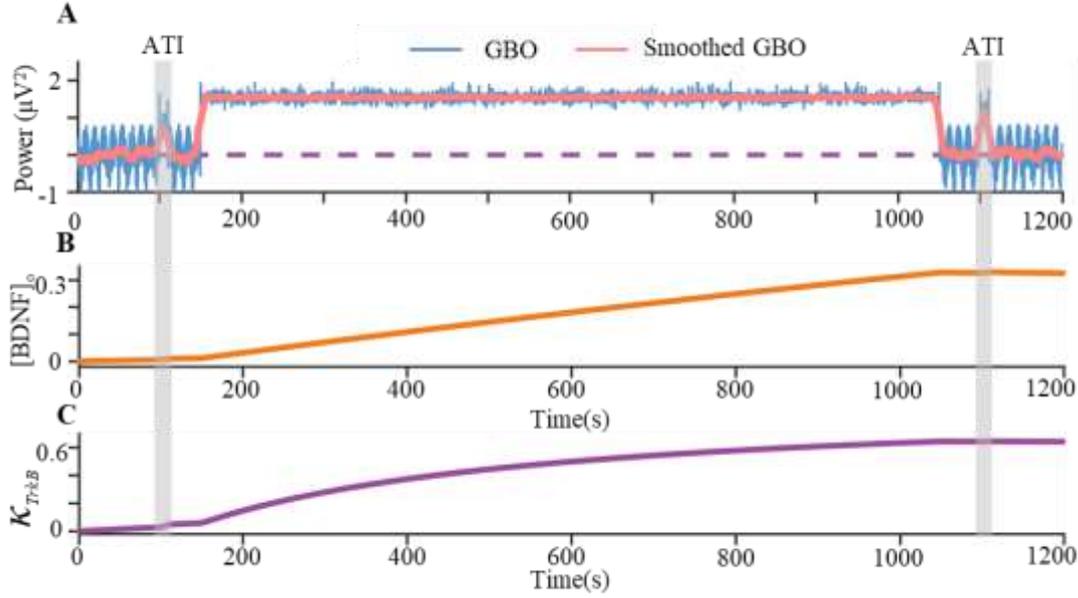

**Figure 5** The dynamical responses of the GBO power, BDNF and the network connection ratio under tDCS. A depicts the GBO diagram of neuron population discharge; B and C illustrate the variation in BDNF concentration and the network connection ratio throughout the entire process.

In order to find out how the changes of tDCS parameter such as the single stimulation period duration $\tau_{tDCS}$ affects network connection ratio $\kappa_{BDNF}$, we calculated the effects of different stimulation $\tau_{tDCS}$ on $\kappa_{BDNF}$. By adjusting $\tau_{tDCS}$ from 0.1 minutes to 30 minutes, the simulation results of the model indicate that the longer $\tau_{tDCS}$ is, the higher the $\kappa_{BDNF}$ will be, as shown in **Figure 6**. It can be seen that with increasing stimulation time, $\kappa_{BDNF}$ shows an approximately linear growth trend. In practical applications, the $\tau_{tDCS}$ can reach up to 20-40 minutes[43], which could be found in our results that can induce an improved brain network connection increase more than 50%.

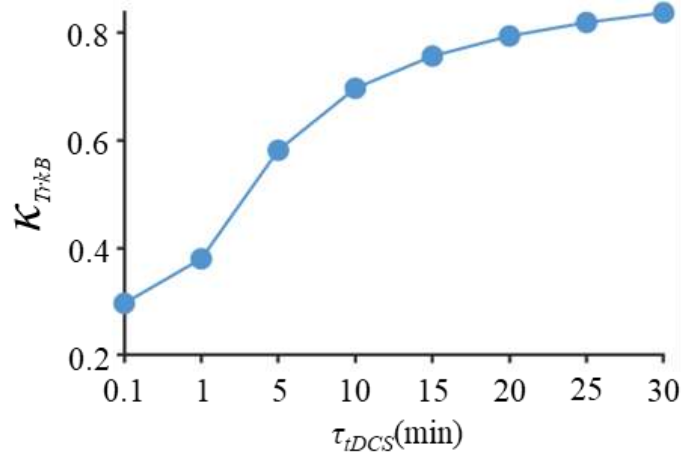

**Figure 6**: The effect of tDCS stimulation with different durations $\tau_{tDCS}$ on the network connection $\kappa_{BDNF}$.

### 3.2 Neuroplasticity and Gamma Oscillation under rTMS Stimulation

Similarly, to explore the cortical neuronal discharge response under rTMS, we conducted two short ATI stimuli before and after a long (approximately 15 minutes) rTMS session. We found that, similar to tDCS stimulation, both the synchronization and discharge frequency of neuron population discharge were enhanced, as shown in **Figures 7 - 8**. **Figures 8A-C** represent three time intervals from **Figure 7**: the resting state before the first ATI stimulation and detailed views of the two ATI phases. **Figures 8D-F** display the average power spectral distribution of the corresponding neuron population discharge at left panel in **Figures 8A-C**.

From the spiking graphs of neuron population discharge in **Figure 7**, it can be observed that, similar to tDCS stimulation, the discharge alternated between sparse and dense patterns. During sparse discharge, the model exhibited a low-frequency discharge state, while during dense discharge, it displayed a high-frequency, synchronized discharge state. Moreover, during the second ATI phase after rTMS stimulation (**Figure 8C**), the synchrony was higher compared to the first ATI phase (**Figure 8B**). Additionally, from the corresponding power spectral graphs (**Figures 8E - 8F**), it can be seen that the gamma- band power during the second ATI phase (**Figure 8**F) was stronger than that during the first ATI phase.

From **Figure 8**, we can conclude that continuous rTMS stimulation for 15 minutes induced changes in the network connetions, affecting the results before and after rTMS stimulation. The synchrony of the model was enhanced, accompanied by an increase in the gamma -band power.

In previous research, Koch et al. [12] found that TMS can enhance local gamma oscillations by altering default network patterns, indicating enhanced synchronization of brain discharge after long-term training. Reinhart and Nguyen's work[37] suggests that synchronization of brain waves can improve working memory in older adults and younger individuals with poor brain function, potentially improving cognitive abilities in populations with poor cognitive function. Asynchronous activity in the frontal cortex may impair adaptive control and learning[38].

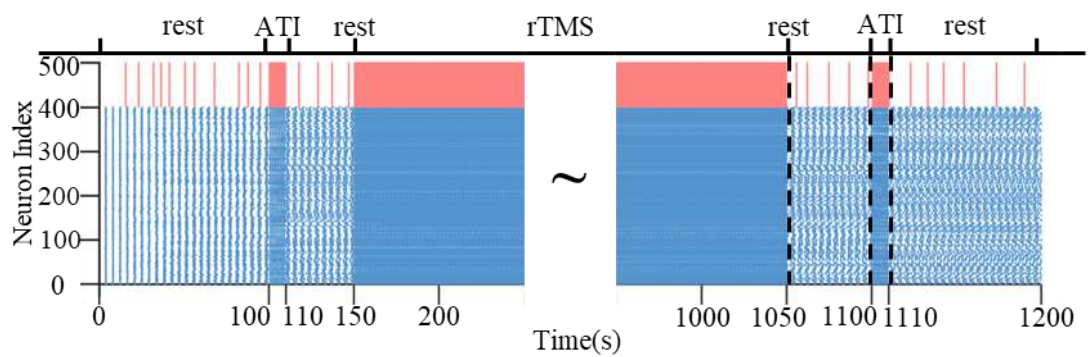

Figure 7 Spiking graph of neuron population discharge during different phases of rest, ATI, and rTMS stimulation phases.

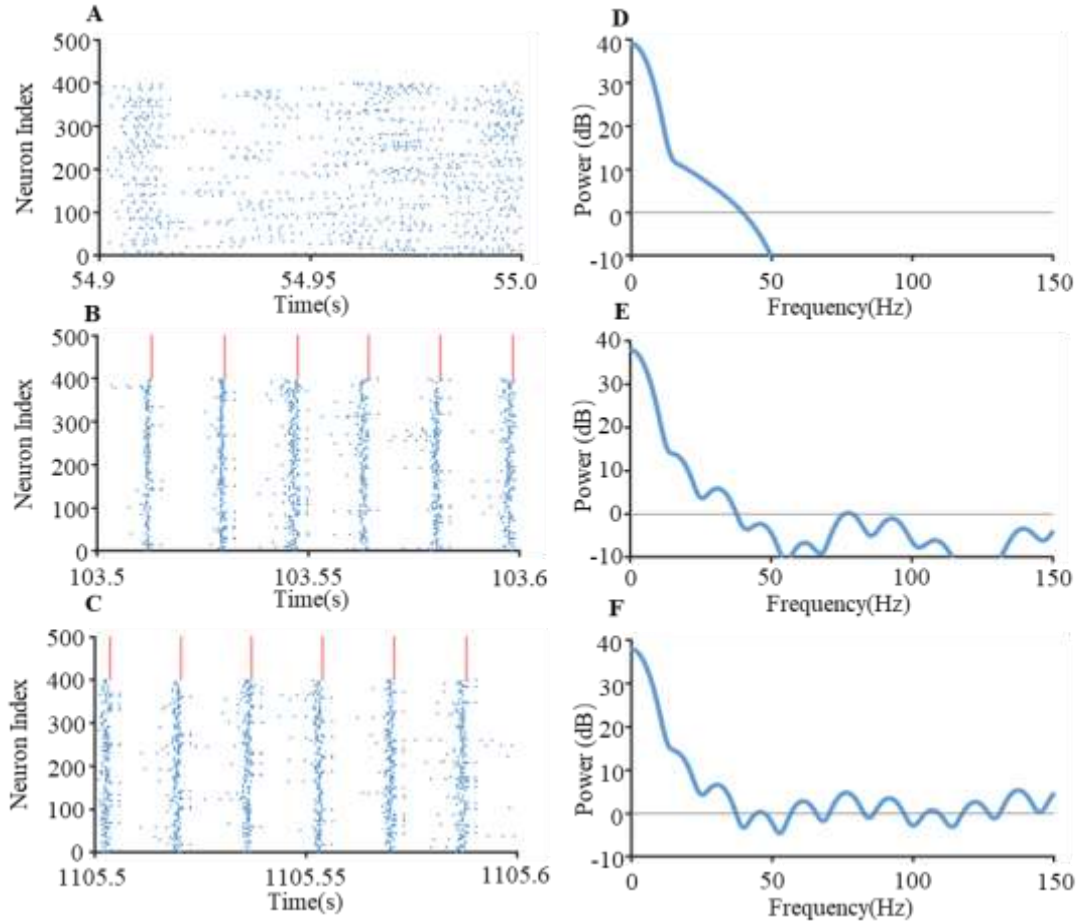

**Figure 8** Temporal and frequency responses of neuron population discharge during stimulation phases selected from **Figure 7**. A-C are detailed spiking diagrams from the time domain of **Figure 7**: the first rest phase (55.5-55.6 seconds), the first ATI phase (103.5-103.6 seconds), and the second ATI phase (1105.5-1105.6 seconds). **D-F** represent the power spectral diagrams of neuron population discharge during the corresponding phases to the left panel, respectively.

Similar to tDCS, rTMS stimulation can induce changes in BDNF concentration, thereby enhancing connectivity strength and neuronal discharge frequency. **Figure 9** depicts the time evolution of average power of GBO of neuron population, along with the corresponding trend of BDNF and $\kappa_{BDNF}$ over time. In the resting state (see **Figure 8D**), the model hardly exhibits neuronal discharge with gamma-band frequency. However, during two ATI phases and the rTMS stimulation phase, the model produces gamma-band discharge with frequency higher than that of the resting state, accompanied by noticeable changes in BDNF concentration (**Figure 9B**) and the network connection ratio $\kappa_{BDNF}$ (**Figure 9**C). During the two ATI phases, it's evident that the GBO response and synchronization of neuron population discharge arise mainly from the varied BDNF concentrations between the two ATI phases.

From these results, we can conclude that continuous rTMS stimulation enhances the GBO of neuron population, which, in turn, promotes BDNF increase. Furthermore, the increase in BDNF drives improvements in structural connections and plasticity. The fluctuation in BDNF concentration and $\kappa_{BDNF}$ reflect increased connectivity strength in the E-I balanced network and, consequently, external manifestations of brain plasticity after the rTMS therapy.

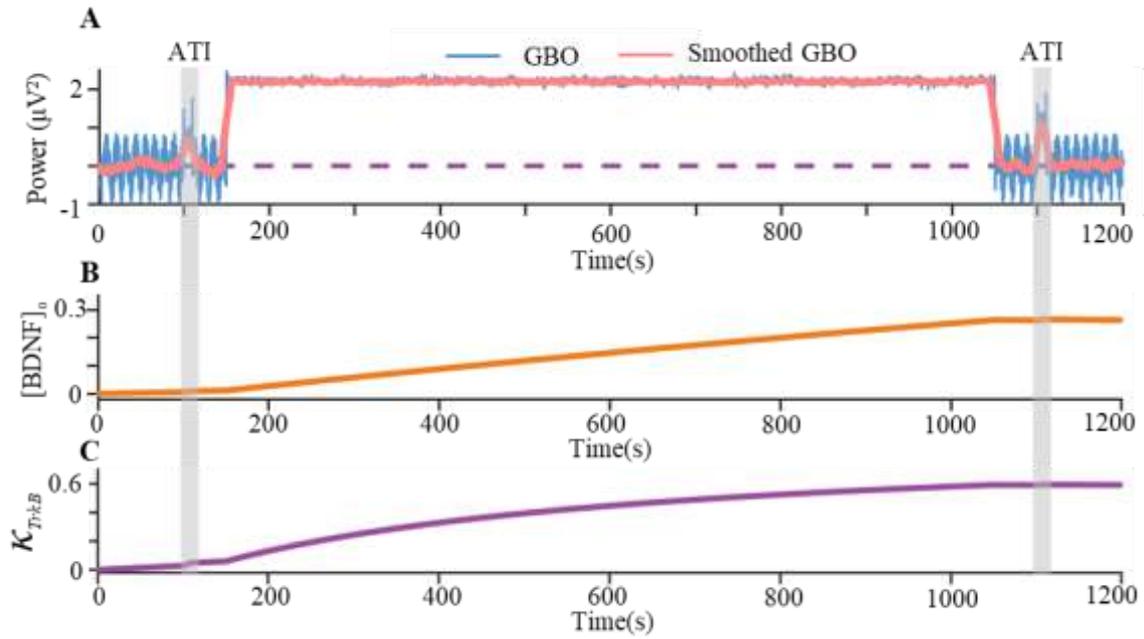

**Figure 9** The dynamical responses of the GBO power, BDNF and the network connection ratio under rTMS. A depicts the GBO diagram of neuron population discharge; B and C illustrate the variation in BDNF concentration and the network connection ratio throughout the entire process.

To unfold the influence of different stimulation frequencies of rTMS on the discharge response of neuron populations under the same stimulation intensity, we varied the stimulation frequency from 1Hz to 100Hz. The simulation results indicate that the model only generates gamma responses when the input stimulation frequency of rTMS increases to above 5Hz, as shown in **Figure 10**. Stimulation parameters higher than 5Hz correspond to the commonly used clinical form of rTMS stimulation, as clinical studies suggest that stimulation below 5Hz may induce inhibition[44].

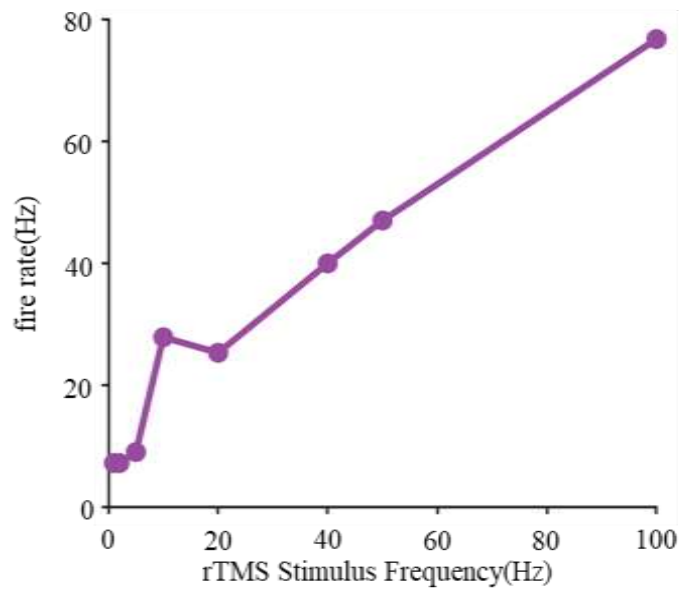

**Figure 10** Firing rate response of neuron population for the changes of rTMS frequency.

**3.3 Gamma oscillations drive the reshaping of neural networks**

To validate the relationship between external stimulus intensity, mean power of gamma band during the long-lasting stimulation, and the increase of neuronal network connection ratio $\Delta\kappa_{BDNF}$ between the two ATI phases. The results are shown in **Figure 11**.

**Figure 11**A describes the changes in averaged firing rate of neuron population by increasing stimulation intensities of tDCS and rTMS. We applied different intensities of stimulation to the model, with intensities ranging from 0 to 20 in increments of 0.5, and observed that the averaged firing rate increased with the increase of stimulation intensities $A_{tDCS}/A_{rTMS}$. Also, it can be seen that when $A_{tDCS}$ and $A_{rTMS}$ reaches a certain value (tDCS stimulation greater than 1.5, rTMS stimulation greater than 9), the firing rate enter the gamma- band domain, and the growth rate of tDCS stimulation (blue dots) is significantly higher than that of rTMS stimulation (orange dots). Experimental results also indicate that transcranial stimulation require a certain level of stimulation intensity[45].

In **Figure 11**B, by gradually increasing the stimulus intensities of tDCS/rTMS, we observed that $\Delta\kappa_{BDNF}$ increased in a same way with the GBO in **Figure 11**A. That means the relations between GBO and $\Delta\kappa_{BDNF}$ reaches linear, which is shown **Figure 11**C. The results above indicate that $\Delta\kappa_{BDNF}$ which represent the improvement level of neuronal network connection induced by tDCS/rTMS, is highly dependent on the presence of GBO state for the neuron population. From a modeling perspective, it suggests the inherent mechanism of GBO in the brain connection improvement by the external stimulation treatment.

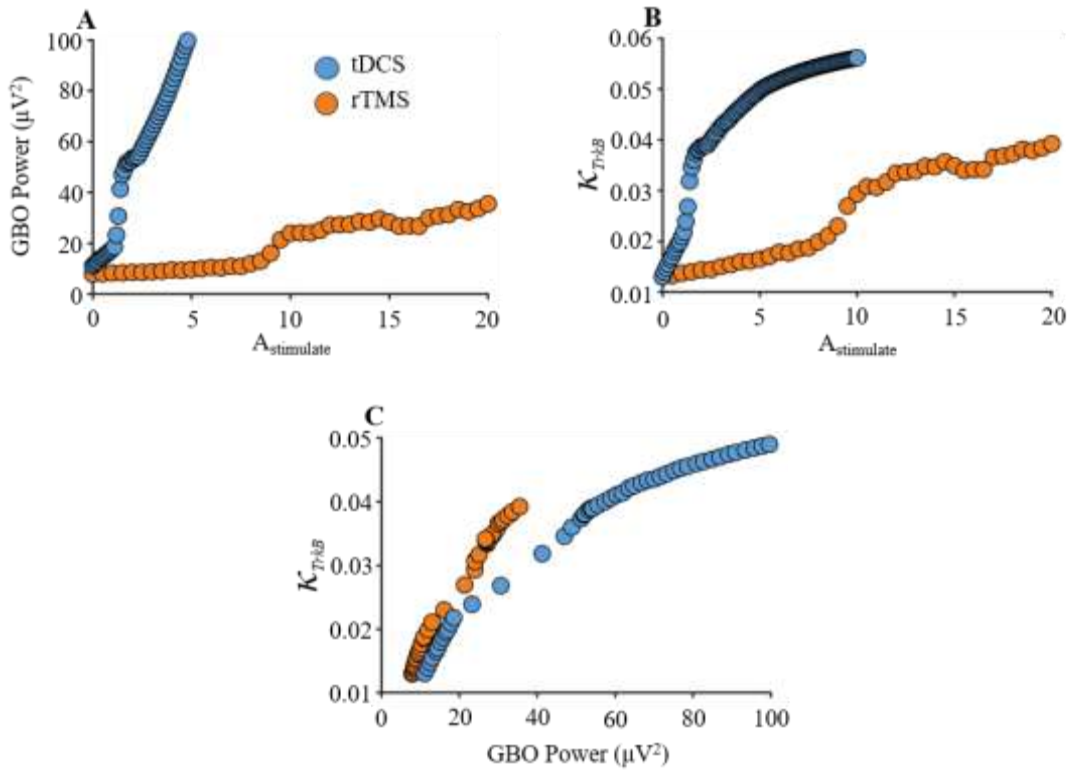

**Figure 11** Gamma oscillation modulated the brain network connection re-plasticity. A shows the mean of the sustained GBO power during the tDCS(blue) or rTMS(orange) with different stimulation amplitude; B represents the increase of the network connection ratio, kBDNF after the stimulation than the before; C illustrates the correlation distribution between GBO and $\Delta\kappa_{BDNF}$.

## 3.4 Brain network correlation and gamma oscillation Encoding based on tDCS-driven EEG data

To validate the simulation results of model, we selected publicly available data from participants undergoing transcranial stimulation therapy, sourced from the OpenNeuro dataset[46]. These data consisted of EEG recordings of participants performing attention tasks before, during, and after tDCS, with each participant undergoing one or two sessions spaced a week apart. Each recording lasted approximately 20 minutes, and the data were sampled at 512Hz with 64 channels. We chose 20 representative trials from the attention experiment data before and after transcranial treatment and compared the corresponding gamma-band responses in cognitive experiments. An analysis of the experimental data revealed a significant enhancement in the power spectrum response of gamma band and brain functional connection (FC) states during cognitive experiments after receiving transcranial treatment, compared to before treatment. This is consistent with the results demonstrated by our model.

Firstly, we applied Fourier transform algorithm to transform the original EEG signal from to the frequency domain, obtaining the power spectrum. Meanwhile, after averaging the selected 20 samples of EEG data, we used a bandpass filter method explained in the Model and Methods section to obtain the dynamical trends of the gamma band (30-100Hz) power based on the original data, as shown in **Figure 12**.

In **Figure 12** A, we observed a significant enhancement in the brain's gamma band response (30-100Hz) after receiving transcranial stimulation, with less noticeable enhancement in other frequency bands (<30Hz).

In **Figure 12**B, we found a significant enhancement in gamma responses during attention tasks after receiving transcranial stimulation. This is similar to the results of our model (**Figure 4** E, F), indicating that continuous tDCS stimulation plays a role in enhancing gamma oscillations and improving cognitive levels in the brain.

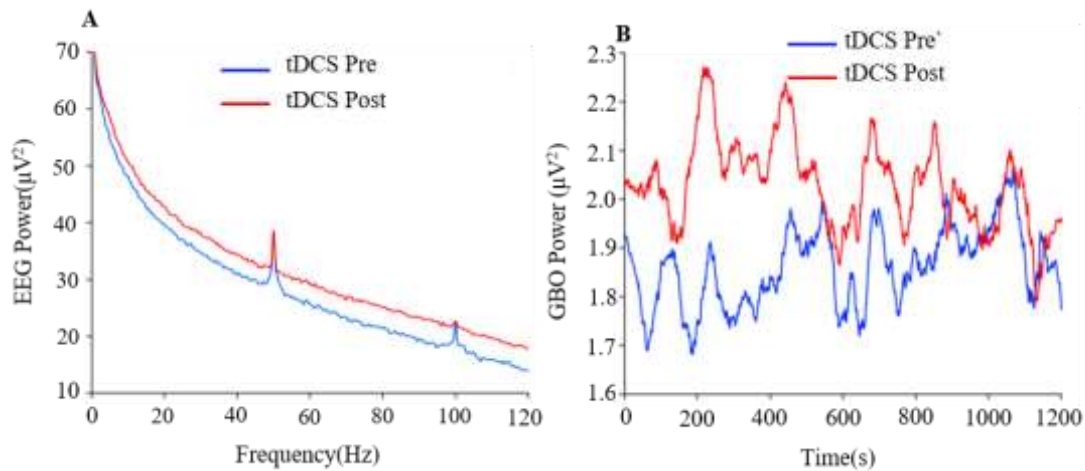

Figure 12 Gamma response of the EEG before and after the tDCS treatment. A shows the power spectrum of EEG from participants subjected to the attentional visual stimuli before and after tDCS. B illustrates the corresponding dynamical responses of gamma-band power.

To further investigate changes of brain connection affected by tDCS, similar discussion in the model sections, we utilized Pearson correlation analysis to examine the whole brain FC based on all electrode EEGs, as well as he directly FC (DFC) between the brain primary visual cortex and the other regions, presented in the form of brain topography maps and histograms, as shown in **Figures**

13-14.

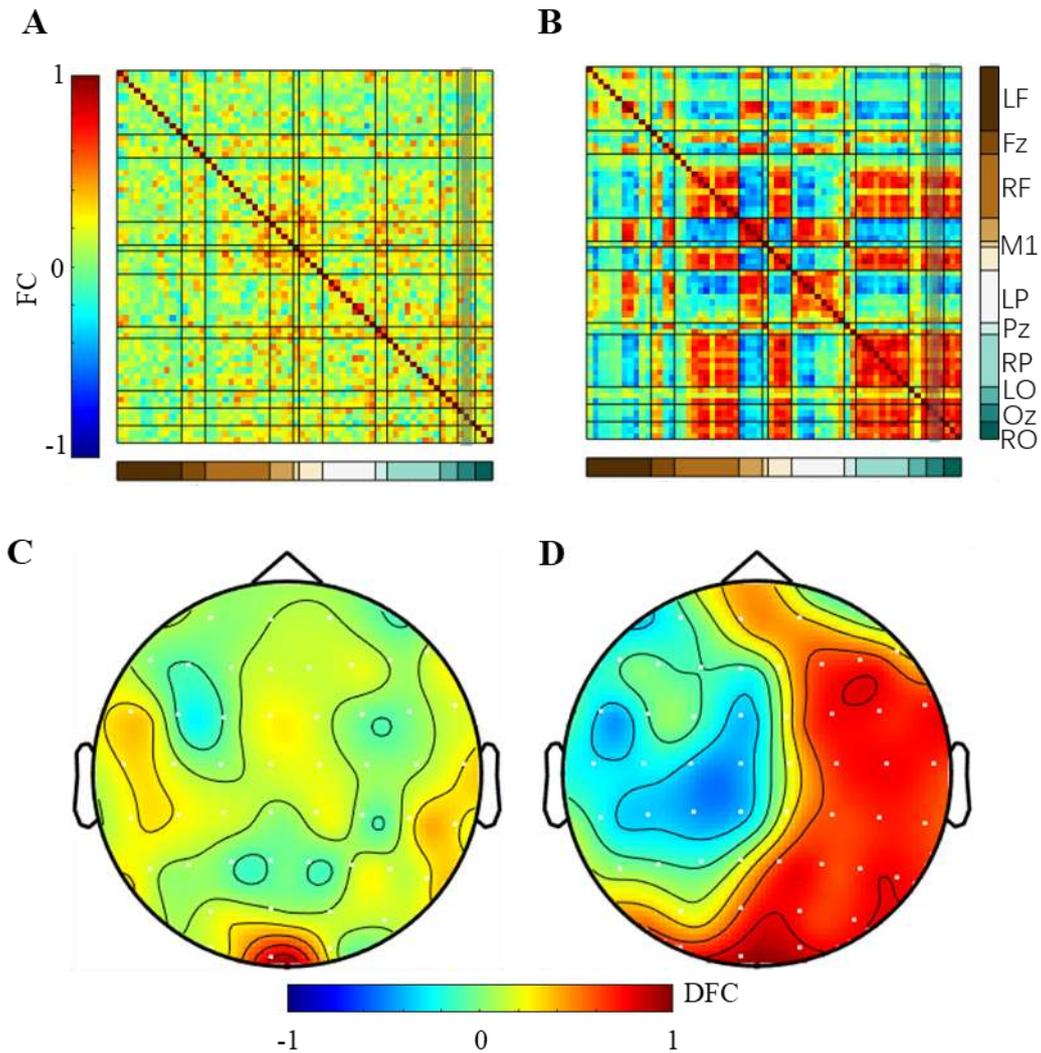

**Figure 13** The increased functional connectivity of the EEG electrodes due to the tDCS stimulation. A-B illustrates the FC distribution of the entire brain during the execution of cognitive stimuli before and after the subjects received tDCS; C-D illustrates the DFC describing the FC of primary visual cortex with the entire brain during the execution of cognitive stimuli before and after the subjects received tDCS.

From **Figure 13**A-B, it is evident that the connection of various brain regions during cognitive tasks is significantly better than that before transcranial stimulation, particularly the enhanced DFC between the occipital primary visual area and the other brain regions (Figures 13C-D). The histogram in **Figure 14** have emphasized that the DFCs in regions of PCL_R, CPL_R, PL_R POL_R (termed in sheet 3) are particularly enhanced by the tDCS. This aligns with the results in our model, where the network connection and gamma oscillation are both enhanced after transcranial stimulation. According to the reference[48], this correlation between the network connection and gamma oscillation signifies the normal perceptual process of the human brain to external stimuli. The connection improvement phenomenon between visual brain regions and the frontal and visual association brain regions (temporal and parietal cortices) has also been observed in other experiments under visual input[49]. Reference[50] suggests that the enhancement of brain functional connectivity actually reflects an increase in synaptic structural connectivity. This corresponds to the enhancement of BDNF and the network connection ratio shown in **Figure 5B**, reflecting increased structural connectivity after tDCS stimulation. The experimental analysis results here indicate an improvement in functional connectivity after tDCS, demonstrating consistency across different methods.

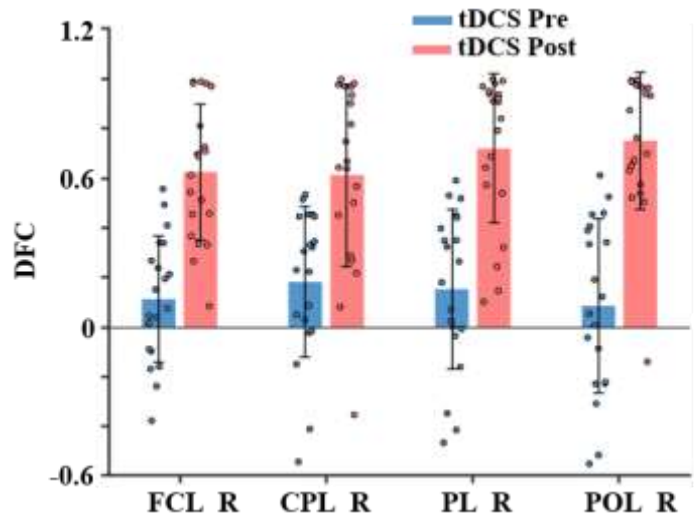

**Figure 14** The histogram of DFC between the primary visual cortex with the region of interests shown below in the Table 3.

**Table 3** The explanation of the abbreviation of regions of interests in **Figure 14**

| abbreviation | The full name of region of interests |
| --- | --- |
| FCL_R | The right side of frontal and center lobe |
| CPL_R | The right side of center and parietal lobe |
| PL_R | The right side of parietal lobe |
| POL_R | The right side of parietal and occipital lobe |

## 4 Discussion

Understanding the brain's neurorepair process under external intervention has been highlighted in the field of rehabilitation of brain diseases. This paper employed a combination of dynamic modeling and data analysis methods. An E-I balanced neural network model was established, incorporating sustained external transcranial stimulations and sensory training input. Additionally, a linear dynamic model of BDNF was constructed based on experimental data. This study investigated the gamma oscillation responses and network connectivity patterns of the neural network before and after sustained external stimulations. The results indicated a significant enhancement in gamma oscillations in the neural network after all the types of stimulations, accompanied by a notable increase in BDNF concentration levels. EEG data analysis following tDCS similarly revealed a marked increase in post-stimulation gamma oscillations, along with enhanced functional connectivity between the primary visual cortex and parietal lobe, and frontal lobe networks.

The results in this paper have illustrated the coding of gamma oscillations at different time scales. Both the model and experimental results indicate that gamma oscillations induced by attention tests after the sustained tDCS, so does cognitive performance is in the experiment. Similar conclusions can be found in literature, including our study, suggesting that the transient steady-state response of gamma oscillations can serve as a biomarker for attention and cognitive levels[16,51]. In the view of the long time scale, the results of this paper suggest that sustained gamma oscillations act as a "driving force" for increasing dendritic spine density induced by the elevation of BDNF. Numerous experiments from a macro perspective have demonstrated that rehabilitation intervention, such as transcranial stimulations, induce gamma oscillations in the brain and enhance cognitive levels[13]. This paper further unfolds, from a multi-scale model perspective, the gamma coding of cognitive improvement in the brain through the external stimulation. [51]

In this article, we discovered that the structural connections of neural networks are significantly enhanced under the influence of BDNF after sustained stimulation through a dynamic modeling approach. The experimental analysis results also indicate an enhancement in functional connections between brain regions. The study by Meng X[52] has proposed that enhanced structural network connections promote the connections of functional network, demonstrating a close relationship between microstructural connections and macroscopic functional connections. Therefore, the integration of the model and experimental results in this article is consistent to some extent, both demonstrating that, under sustained external stimulations, there is a noticeable improvement in both macroscopic functional connections and microstructural connections in the brain. This indirectly validates the intrinsic mechanisms of neurorepair in the human brain.

Non-invasive stimulations, especially transcranial stimulations, have been a highly promising therapeutic approach for cognitive disorders in clinical trials. However, the underlying mechanisms of these external interventions remain unclear. This article, through the combined analysis of a model and the experimental data, provides a close dynamic coupling relationship between external stimulation and gamma oscillations in the brain. This relationship can serve as a crucial theoretical reference for clinical research, such as transcranial stimulation, on how to utilize the gamma oscillation levels for monitoring stimulation.

## Acknowledgement

This work was supported by the National Natural Science Foundation of China (Grant Nos. 12002251). We thank the Leon and his colleagues for provide publicly available EEG data for us accessible to validate the simulation results.

## Data Availability Statement

A portion or all data, models, and code generated or used during the study are available from the corresponding author upon request.

## Conflict of Interest

The authors declare that they have no conflict of interest.